\documentclass[11pt,manuscript]{aastex}

\shorttitle{Hinode/XRT and RHESSI hot plasma}
\shortauthors{Reale et al.}

\begin{document}

\title{Comparison of Hinode/XRT and RHESSI detection of hot plasma in the non-flaring solar corona}

\author{Fabio Reale\altaffilmark{1}}
\affil{Dipartimento di Scienze Fisiche \& Astronomiche, Universit\`a di
       Palermo, Sezione di Astronomia, Piazza del Parlamento 1, 90134 Palermo,
       Italy}
\author{James M. McTiernan}
\affil{Space Sciences Laboratory, University of California, Berkeley, CA 94720-7450, USA}
\author{Paola Testa}
\affil{Harvard-Smithsonian Center for Astrophysics, Cambridge, MA 02138, USA}

\altaffiltext{1}{also INAF - Osservatorio Astronomico di Palermo ``G.S.
       Vaiana'', Piazza del Parlamento 1, 90134 Palermo, Italy}

\begin{abstract}
We compare observations of the non-flaring solar corona made simultaneously with Hinode/XRT and with RHESSI. The analyzed corona is dominated by a single active region on 12 November 2006. The comparison is made on emission measures. We derive emission measure distributions vs temperature of the entire active region from multifilter XRT data. We check the compatibility with the total emission measure values estimated from the flux measured with RHESSI if the emission come from isothermal plasma. We find that RHESSI and XRT data analyses consistently point to the presence of a minor emission measure component peaking at $\log T \sim 6.8-6.9$. The discrepancy between XRT and RHESSI results is within a factor of a few and indicates an acceptable level of cross-consistency.
\end{abstract}


\section{Introduction}

Nanoflares in multistranded loops are among the best candidate to explain the heating of the confined solar corona. 
The existence of nanoflares is still under debate,
and a strong evidence in support of nanoflares would be the
detection of $\sim 10$ MK plasma \citep{1995itsa.conf...17C} in the quiescent corona.
Recent observations suggest that such hot plasma may indeed be
present at low levels in nonflaring active regions
\citep{2006SoSyR..40..272Z,2007AstL...33..396U,2009ApJ...696..760P,2009ApJ...697...94M,2009ApJ...693L.131S}.

The {\it X-Ray Telescope} (XRT, \cite{2007SoPh..243...63G}) on board the Hinode mission \citep{2007SoPh..243....3K} is sensitive in the energy band from $\sim 0.15$ to more than 3 keV  and can detect emission from plasma with temperatures from $\sim 1$ to several tens MK. Multi-filter, high-sensitivity and high resolution observations made with the XRT have shown evidence that hot plasma may be indeed widespread in active regions and provide even more support to the nanoflare scenario \citep{2009ApJ...698..756R}. 

Such hot plasma component appears to be anyhow a minor contribution to the overall budget of the X-ray emitting quiescent corona and therefore the evidence may be easily affected by systematic errors, and will continue to need further support from independent data and analysis. An important testing ground becomes the cross-check with other instruments able to detect such hot component.

The RHESSI spacecraft can observe solar X-rays and $\gamma$-rays in the
energy range from 3 keV to approximately 17 MeV and can detect emission from plasmas
with temperatures approximately as low as 5 MK up. It has been designed especially to observe flares and therefore its sensitivity to the quiescent corona is limited both because of the low flux and of the temperatures close to the lower boundary. A recent systematic analysis of RHESSI data has detected the usual presence of high-temperature solar emission
\citep{2009ApJ...697...94M}. 

XRT and RHESSI have overlapping energy bands just in the range of interest and provide an important opportunity to test further the hot tail of the plasma emitting in the soft X-ray band.
In the present work, we compare the emission detected with RHESSI and that detected with Hinode/XRT from the Sun on 12 November 2006 and check for compatibility.

\section{Data analysis}
\subsection{XRT Data}
\label{sec:xrt}

\begin{figure}[!t]
  \centering \includegraphics[width=6cm]{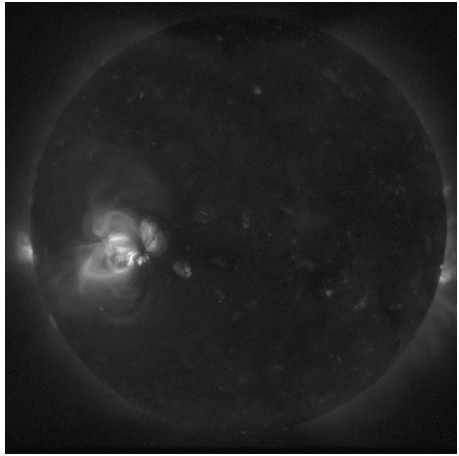} 
  \caption{The solar corona as imaged by Hinode/XRT on 12 November 2006 in the Al\_poly filter (the grey scale is [DN/s]$^{0.3}$ between 0 and 5000 DN/s). AR10923 dominates the X-ray emission.  
 }
  \label{fig:sun}
\end{figure}

We consider the same XRT data as in  \cite{2009ApJ...698..756R} (see also \cite{2007Sci...318.1582R}) and take their results as basic for the present work. The
field of view (512$\times$512 arcsec$^2$) includes an active region
(AR10923) observed close to the Sun center on 12 November 2006. In the following we will assume that the active region is the dominant contributor to the X-ray emission at that time (Fig.~\ref{fig:sun}). The filters used were Al\_poly (F1), C\_poly (F2), Be\_thin (F3),
Be\_med (F4) and Al\_med (F5), with exposure times of 0.26~s,
0.36~s, 1.44~s, 8.19~s and 16.38~s, respectively.  F1 and F2 are sensitive mostly in the 0.2-3 keV energy band, F3 in the 0.6-3 keV band, F4 in the 0.8-3 keV band, F5 in the 0.8-2 keV band \citep{2007SoPh..243...63G}. The selected dataset covers one hour, starting at 13:00 UT, and the time interval
between one exposure and the next in the same filter is about five
minutes (12 images in each filter). The images were averaged over the whole hour, to improve for signal-to-noise ratio, and co-aligned with a cross-correlation technique.

The analysis in \cite{2009ApJ...698..756R} derived information about the global thermal structure of the active region both spatially resolved and along the line of sight. This is obtained by combining the information coming from the data in all filters. In particular, temperature and emission measure maps, i.e. single values for each pixel, are obtained for a given filter ratio. This is done for several filter ratios. Each filter ratio samples the thermal information in a slightly different way and therefore, in the end, a multiple, although limited, sampling of the emission measure distribution along the line of sight is available for each pixel. The hardest filter ratio F4/F5 gives information on the hot component of the emission and the soft filter ratios (e.g. F5/F1) on the cooler component.
Then it has been summed up the emission measures of all pixels falling in the same temperature bins, building an emission measure distribution vs temperature for each considered filter ratio.
The next steps were aimed at deriving the underlying ``parent" emission measure, structured also along the line of sight, able to yield the observed weight-filtered distributions. The task was accomplished by means of Monte Carlo simulations: it was assumed a basic simple parent emission measure distribution along the line of sight, typically made by two isothermal components for each pixel; this parent distribution was randomized to account for variations from pixel to pixel; from the resulting randomized distribution the emission expected in all filters was computed; the emission values were all randomized according to Poisson statistics. In the end, a fake image for each filter was obtained and analyzed as done for the real data, to find global emission measure distributions to be compared to the those obtained from the observation. All this procedure was repeated several times until the simulated EM(T)'s were similar to the ``observed" ones.

\begin{figure}[!t]
  \centering \includegraphics[width=6cm]{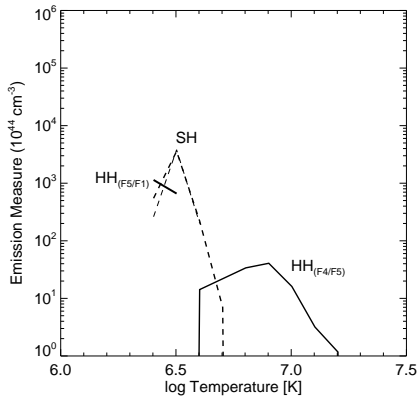} 
  \caption{Emission measure distributions vs temperatures measured with two XRT filter ratios  (a soft ratio, F5/F1, and the hardest ratio, F4/F5) for two subregions of the active region: a hard-hot (HH) subregion ({\it solid lines}) and a soft-hot (SH) subregion ({\it dashed lines}). In subregion HH  a hot component ($\log T \sim 7$) is measured with the hard filter ratio well separated from the cooler component measured with the soft filter ratio (as labelled in parenthesis). In the SH subregion both filter ratios measure only one very similar cool component (thinner line for F5/F1), hotter than the cool HH component (see also Fig.5 in \cite{2009ApJ...698..756R}).  
 }
  \label{fig:EMT_XRT}
\end{figure}

The analysis was applied in particular to two subregions showing homogeneous properties, a bright one, named SH (``Soft-Hot''), where the temperature is relatively higher than the surroundings in the soft filter ratios, and a fainter one, named HH (``Hard-Hot"), where the temperature is higher in the hardest filter ratio F4/F5, as shown in Fig.~\ref{fig:EMT_XRT}. As a final result, it was shown that a bimodal parent emission measure distribution is able to describe both subregions. The difference between the regions can be consistently explained with a temperature shift of the cool and high emission measure component. In region SH, the cool component is shifted to higher temperature, where the hard filter ratio is able to detect it. This does not occur in region HH, in which the hard filter ratio is more sensitive to the hot and small component, that instead can remain practically unchanged. 

Our analysis here takes the distinct but similar parent emission measures obtained for regions SH and HH as starting points. Our final target is to check the compatibility of the temperature structure obtained from Hinode/XRT data with the one obtained from RHESSI. To accomplish this, we need to determine the emission measure distribution of the entire active region. We have found it very difficult to obtain this goal with the same approach as pursued separately for regions HH and SH, i.e. with a Monte Carlo simulation randomizing a single parent EM(T) distribution. This simply means that the description of the entire active region needs a more complex randomization pattern. We have realized that, for our purposes, it is enough to extrapolate the results obtained for the two subregions. We have then decided to scale independently the two parent EM(T) distributions, and to find for each of them the best possible match between the related output EM(T) distributions filtered through the filter responses and the ones derived from the observation for the entire active region. The advantage of this approach is also that the difference between the two scaled parent distributions provides an order of the overall error. The scaling has been performed simply with a single multiplication factor on the whole parent EM(T). The goodness of the matching has been measured differently for the two parent EM(T) distribution, due to the qualitative difference between the output EM(T)'s.

More specifically, we have scaled the parent EM(T) of region HH so as that the output EM(T) obtained from the hardest filter ratio F4/F5 matches the hot part of the corresponding F4/F5 output EM(T) obtained for the entire active region (see Fig.~\ref{fig:RHE_XRT}a). We obtain a factor 5 with a sensitivity of 0.5. The output EM(T)'s of region SH lack completely the hot component. Then, we have scaled the parent EM(T) of region SH so as that the total output EM(T) obtained from one of the soft filter ratios matches the total output EM(T) obtained for the entire active region. With a factor 6.5 (with a sensitivity of 0.5) the total EM(T)'s differ by less than 5\% for both ratios F5/F1 and F4/F1. 

We take these two scaled parent EM(T)'s for comparison with results obtained from RHESSI. 
We point out that ours is not a proper differential emission measure (DEM) reconstruction, rather a forward modeling in which we use simplified EM(T) functions. This is anyway a sensible approach also considering that the temperature resolution of imaging instruments such as XRT is intrinsically limited and not able to constrain the very fine details of the DEM. For the sake of clarity, in the figures we have adopted the convention to mark any XRT observed EM(T) with a line and any parent EM(T) with a histogram.

\subsection{RHESSI Data}
\label{sec:rhessi}

%- RHESSI data:
% Added some description, 22-jun-2009, jmm
The RHESSI spacecraft, launched in February 2002, carries nine germanium detectors which are used to observe solar X-rays and $\gamma$-rays in the energy range from 3~keV to approximately 17~MeV, with better than 1~keV FWHM energy resolution \citep{2002SoPh..210....3L,2002SoPh..210...33S}. Using RHESSI data, \cite{2009ApJ...697...94M} has found that emission with temperatures from 6 to 10~MK are typically present during active times, in the absence of solar flares.

The RHESSI count rate has been measured for five of the detectors (numbers 1, 3, 4, 6, 9) for the time period 1255 - 1300 UT for 12-nov-2006. The non-solar background has been determined from the RHESSI spacecraft position and subtracted as discussed by \cite{2009ApJ...697...94M}. After background subtraction, the RHESSI temperature measurement for this time period gives a value of $T_{RHE} = 8.1$ MK and an emission measure $EM_{RHE} = 4.1 \times 10^{45}$ cm$^{-3}$. The RHESSI detectors used in the calculation have an energy range from 3 to 300~keV. There were excess counts in the energy range from 4~keV to 10~keV for this time interval. The temperature determination is made uncertain by the relatively flat spectrum in the energy range used in the calculation, and the relatively low signal-to-noise level; the excess count rate in this range is only approximately 50\% of the background count rate. Therefore, in order to compare the RHESSI and Hinode/XRT results we have put the RHESSI results under a different form.

From the flux measured with RHESSI, we have derived the emission measure value as if the emission is entirely from an isothermal plasma volume. We have done this for temperatures $6.1 \leq \log T \leq 7.1$. This is a temperature range that is appropriate for the XRT, but only overlaps partly with the temperature range of RHESSI observations ($\log T \geq 6.8$). 

\subsection{Comparison RHESSI--XRT }

\begin{figure}[!t]
  \centering \includegraphics[width=6cm]{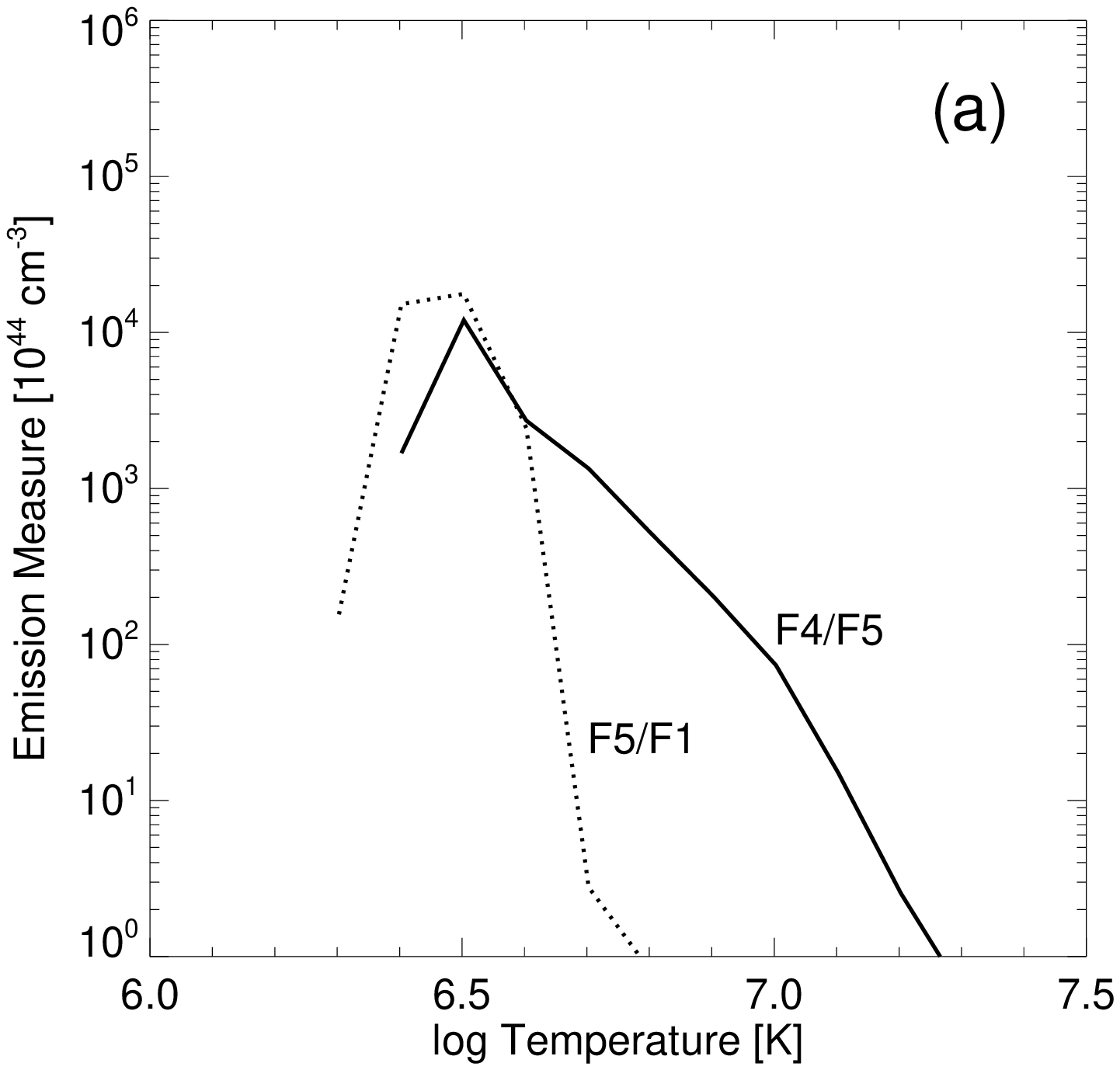} 
  \centering \includegraphics[width=6cm]{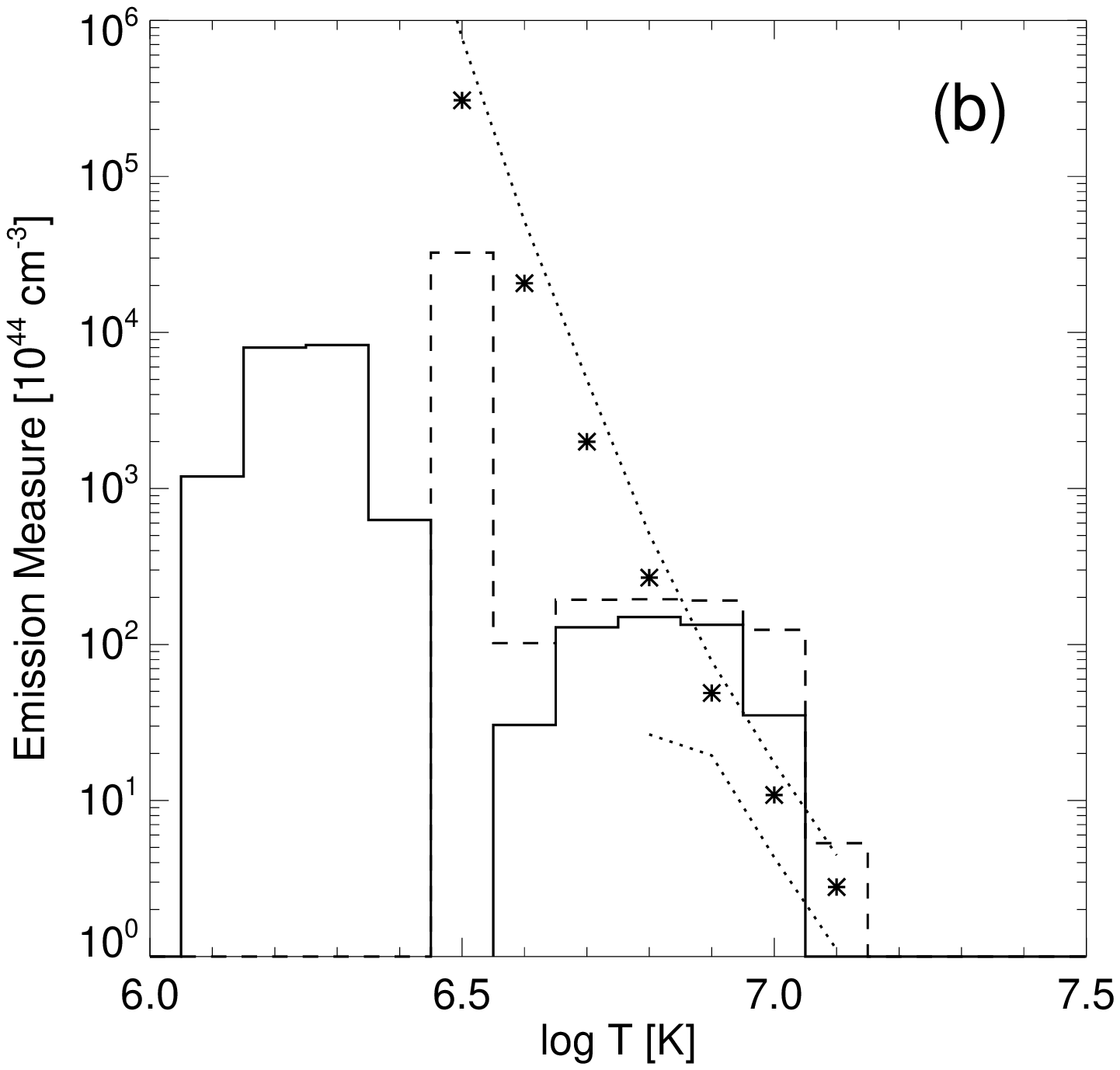} 
  \caption{{\it (a)} Emission measure distributions vs temperature of the entire active region measured with the two labelled filter ratios of Hinode/XRT, i.e. the hard one F4/F5 ({\it solid line}) and the soft one F5/F1 ({\it dotted line}) (see also Fig.3 in \citet{2009ApJ...698..756R}). {\it (b)} Comparison between emission measures obtained from Hinode/XRT and from RHESSI. Parent emission measure distributions vs temperature extrapolated from the XRT hard-hot (HH) subregion ({\it solid histogram}) and the XRT soft-hot (SH) subregion ({\it dashed histogram}) to the entire active region are shown. RHESSI emission measure values ({\it data points}) are obtained assuming isothermal plasma at the temperature of each point. Confidence strips for these values ({\it dotted lines}) are also shown. Note that RHESSI measurements for the temperature range $\log T \leq 6.6$ are off the scale, reflecting the limitations of RHESSI in this low $T$ range.  
 }
  \label{fig:RHE_XRT}
\end{figure}

In Fig.~\ref{fig:RHE_XRT}b we have compared the RHESSI emission measure values as computed in Sec.\ref{sec:rhessi} with the scaled parent EM(T) distributions obtained from Hinode/XRT as in Sec.\ref{sec:xrt}. The histograms in Fig.~\ref{fig:RHE_XRT}b are the parent EM(T) distributions which correspond to the distributions measured with the soft and hard XRT filter ratios in Fig.~\ref{fig:RHE_XRT}a.  

Of course, strictly speaking, in Fig.~\ref{fig:RHE_XRT}b the values related to the XRT are not directly comparable with those related to RHESSI. For the XRT we show temperature-resolved emission measure distributions, while for RHESSI we show integrated values, each of which should be compared with the integrated XRT histograms. We think that the comparison cannot be made more detailed than that, because we are not able to resolve the temperature distribution from RHESSI data alone, and we want to maintain the temperature information of the XRT analysis.

 Nevertheless, in the logarithmic scale we can afford at least a rough but direct comparison. We see that the emission measure values obtained from RHESSI decrease steeply with temperature. For $\log T \leq 6.8$ they clearly become much higher than the level of emission measure obtained from the XRT, at comparable temperature. This is clearly consistent with the fact that RHESSI is hardly sensitive to plasma below that temperature. On the other hand, the plot shows an overall consistency between the RHESSI values and the hot components derived from the analysis of the XRT data. If we group the entire XRT hot component into a single isothermal component we would not be too far from the RHESSI values found at $\log T \approx 6.8-6.9$. For instance, for $\log T = 6.8$, RHESSI yields an emission measure $3 \pm 2 \times 10^{46}$ cm$^{-3}$ which may be compared to the total emission measure 4--7 $\times 10^{46}$ cm$^{-3}$ obtained from the XRT for $\log T \geq 6.7$. If any, the XRT hot component appears overestimated with respect to RHESSI result by a factor of a few. A number of factors may help to explain the mismatch. The difference between the XRT parent distributions already shows a range of variation propagated from their derivation. Unknown and systematic calibration errors might easily explain shifts by $\sim 0.1$ in log T and factors 2-3 in emission measure values. In particular, the calibration of XRT filters is not completely stable yet \citep{2009ApJ...698..756R}, and we are pushing our analysis to the very end of RHESSI sensitivity range, where typically the calibration is less constrained.

\section{Conclusions}

In our opinion, Fig.~\ref{fig:RHE_XRT}b, on the one hand, provides indications about the cross-calibration between the RHESSI and Hinode/XRT and suggests that the two instruments provide overall consistent results, within the limitations of the present analysis. On the other hand, the figure gives a further support to the existence of a relatively hot permament component in the non-flaring solar corona. The details of the plasma thermal distribution are certainly still to be better defined, due to the presence of a number of sources of uncertainties. One major problem arises from the fact that this component results to be intrinsically minor and therefore difficult to detect from the two instruments for different reasons. XRT filters are sensitive to, and therefore dominated by, the cooler and much stronger emission measure component at $\log T \sim 6.3-6.4$. For RHESSI it is instead difficult to detect such low emission measure component at the lower end of its temperature sensitivity range. Although this work goes in the direction of confirming this finding, further independent evidence is required to put a conclusive word.

\bigskip
\acknowledgements{We thank H. Hudson for triggering this work, J. Klimchuk and the anonymous referee for comments and suggestions. Hinode is a Japanese mission developed and launched by ISAS/JAXA, with NAOJ as domestic partner and NASA and STFC (UK) as
international partners. It is operated by these agencies in co-operation
with ESA and NSC (Norway).  FR acknowledges support from
Italian Ministero dell'Universit\`a e Ricerca and Agenzia Spaziale Italiana
(ASI), contracts I/015/07/0 and I/023/09/0. PT is supported by
NASA contract NNM07AA02C to SAO. JM is supported by NASA contract NAS5-98033 and NASA grant NNX08AJ18G. }

\bibliographystyle{apj}
%\bibliography{references}

\begin{thebibliography}{12}
\expandafter\ifx\csname natexlab\endcsname\relax\def\natexlab#1{#1}\fi

\bibitem[{{Cargill}(1995)}]{1995itsa.conf...17C}
{Cargill}, P.~J. 1995, in Infrared tools for solar astrophysics: What's next?,
  ed. J.~R. {Kuhn} \& M.~J. {Penn}, 17--+

\bibitem[{{Golub} {et~al.}(2007){Golub}, {Deluca}, {Austin}, {Bookbinder},
  {Caldwell}, {Cheimets}, {Cirtain}, {Cosmo}, {Reid}, {Sette}, {Weber},
  {Sakao}, {Kano}, {Shibasaki}, {Hara}, {Tsuneta}, {Kumagai}, {Tamura},
  {Shimojo}, {McCracken}, {Carpenter}, {Haight}, {Siler}, {Wright}, {Tucker},
  {Rutledge}, {Barbera}, {Peres}, \& {Varisco}}]{2007SoPh..243...63G}
{Golub}, L. {et~al.} 2007, \solphys, 243, 63

\bibitem[{{Kosugi} {et~al.}(2007){Kosugi}, {Matsuzaki}, {Sakao}, {Shimizu},
  {Sone}, {Tachikawa}, {Hashimoto}, {Minesugi}, {Ohnishi}, {Yamada}, {Tsuneta},
  {Hara}, {Ichimoto}, {Suematsu}, {Shimojo}, {Watanabe}, {Shimada}, {Davis},
  {Hill}, {Owens}, {Title}, {Culhane}, {Harra}, {Doschek}, \&
  {Golub}}]{2007SoPh..243....3K}
{Kosugi}, T. {et~al.} 2007, \solphys, 243, 3

\bibitem[{{Lin} {et~al.}(2002){Lin}, {Dennis}, {Hurford}, {Smith}, {Zehnder},
  {Harvey}, {Curtis}, {Pankow}, {Turin}, {Bester}, {Csillaghy}, {Lewis},
  {Madden}, {van Beek}, {Appleby}, {Raudorf}, {McTiernan}, {Ramaty}, {Schmahl},
  {Schwartz}, {Krucker}, {Abiad}, {Quinn}, {Berg}, {Hashii}, {Sterling},
  {Jackson}, {Pratt}, {Campbell}, {Malone}, {Landis}, {Barrington-Leigh},
  {Slassi-Sennou}, {Cork}, {Clark}, {Amato}, {Orwig}, {Boyle}, {Banks},
  {Shirey}, {Tolbert}, {Zarro}, {Snow}, {Thomsen}, {Henneck}, {McHedlishvili},
  {Ming}, {Fivian}, {Jordan}, {Wanner}, {Crubb}, {Preble}, {Matranga}, {Benz},
  {Hudson}, {Canfield}, {Holman}, {Crannell}, {Kosugi}, {Emslie}, {Vilmer},
  {Brown}, {Johns-Krull}, {Aschwanden}, {Metcalf}, \&
  {Conway}}]{2002SoPh..210....3L}
{Lin}, R.~P. {et~al.} 2002, \solphys, 210, 3

\bibitem[{{McTiernan}(2009)}]{2009ApJ...697...94M}
{McTiernan}, J.~M. 2009, \apj, 697, 94

\bibitem[{{Patsourakos} \& {Klimchuk}(2009)}]{2009ApJ...696..760P}
{Patsourakos}, S., \& {Klimchuk}, J.~A. 2009, \apj, 696, 760

\bibitem[{{Reale} {et~al.}(2007){Reale}, {Parenti}, {Reeves}, {Weber}, {Bobra},
  {Barbera}, {Kano}, {Narukage}, {Shimojo}, {Sakao}, {Peres}, \&
  {Golub}}]{2007Sci...318.1582R}
{Reale}, F. {et~al.} 2007, Science, 318, 1582

\bibitem[{{Reale} {et~al.}(2009){Reale}, {Testa}, {Klimchuk}, \& {Susanna
  Parenti}}]{2009ApJ...698..756R}
{Reale}, F., {Testa}, P., {Klimchuk}, J.~A., \& {Susanna Parenti}. 2009, \apj,
  698, 756

\bibitem[{{Schmelz} {et~al.}(2009){Schmelz}, {Saar}, {DeLuca}, {Golub},
  {Kashyap}, {Weber}, \& {Klimchuk}}]{2009ApJ...693L.131S}
{Schmelz}, J.~T., {Saar}, S.~H., {DeLuca}, E.~E., {Golub}, L., {Kashyap},
  V.~L., {Weber}, M.~A., \& {Klimchuk}, J.~A. 2009, \apjl, 693, L131

\bibitem[{{Smith} {et~al.}(2002){Smith}, {Lin}, {Turin}, {Curtis}, {Primbsch},
  {Campbell}, {Abiad}, {Schroeder}, {Cork}, {Hull}, {Landis}, {Madden},
  {Malone}, {Pehl}, {Raudorf}, {Sangsingkeow}, {Boyle}, {Banks}, {Shirey}, \&
  {Schwartz}}]{2002SoPh..210...33S}
{Smith}, D.~M. {et~al.} 2002, \solphys, 210, 33

\bibitem[{{Urnov} {et~al.}(2007){Urnov}, {Shestov}, {Bogachev}, {Goryaev},
  {Zhitnik}, \& {Kuzin}}]{2007AstL...33..396U}
{Urnov}, A.~M., {Shestov}, S.~V., {Bogachev}, S.~A., {Goryaev}, F.~F.,
  {Zhitnik}, I.~A., \& {Kuzin}, S.~V. 2007, Astronomy Letters, 33, 396

\bibitem[{{Zhitnik} {et~al.}(2006){Zhitnik}, {Kuzin}, {Urnov}, {Bogachev},
  {Goryaev}, \& {Shestov}}]{2006SoSyR..40..272Z}
{Zhitnik}, I.~A., {Kuzin}, S.~V., {Urnov}, A.~M., {Bogachev}, S.~A., {Goryaev},
  F.~F., \& {Shestov}, S.~V. 2006, Solar System Research, 40, 272

\end{thebibliography}

\end{document}